\documentclass[journal=jctcce,layout=twocolumn]{achemso}
\usepackage{amsmath}
\usepackage{amssymb}
\usepackage{color}
\usepackage{graphicx}	
\usepackage{bm}
\usepackage{ulem}

\newcommand{\be}{\begin{equation}}
\newcommand{\ee}{\end{equation}}
\newcommand{\bef}{\begin{figure}}
\newcommand{\eef}{\end{figure}}
\newcommand{\bea}{\begin{eqnarray}}
\newcommand{\eea}{\end{eqnarray}}

\title{Reconstructing the free-energy landscape of Met-enkephalin using 
      dihedral Principal Component Analysis and Well-tempered Metadynamics}

\author{F. Sicard}
\email{francois.sicard@u-bourgogne.fr}
\affiliation{Laboratoire Interdisciplinaire Carnot de Bourgogne, UMR 6303 CNRS-Universit\'e de Bourgogne,
	     9 Avenue A. Savary, BP 47 870, F-21078 Dijon Cedex, France }
\author{P. Senet}
\email{patrick.senet@u-bourgogne.fr}
\affiliation{Laboratoire Interdisciplinaire Carnot de Bourgogne, UMR 6303 CNRS-Universit\'e de Bourgogne,
	     9 Avenue A. Savary, BP 47 870, F-21078 Dijon Cedex, France }
	     
\begin{document}
\maketitle
\begin{abstract}
Well-Tempered Metadynamics (WTmetaD) is an efficient method to enhance the reconstruction of the  
free-energy surface of proteins. WTmetaD guarantees a faster convergence in the long time
 limit in comparison with the standard metadynamics. It still suffers however from the same limitation, 
\textit{i.e.} the non trivial choice of pertinent collective variables (CVs). 
To circumvent this problem, we couple WTmetaD with a set of CVs generated from a dihedral Principal Component Analysis (dPCA) 
on the Ramachadran dihedral angles describing the backbone structure of the protein. The dPCA provides a generic method 
to extract relevant CVs built from internal coordinates. We illustrate the robustness of this method in the case of the small 
and very diffusive Met-enkephalin pentapeptide, and highlight a criterion to limit the number of CVs necessary to biased the 
metadynamics simulation. The free-energy landscape (FEL) of Met-enkephalin built on CVs generated from 
dPCA is found rugged compared with the FEL built on CVs extracted from PCA of the Cartesian coordinates of the atoms.
\end{abstract}
\section{Introduction}
Since the late 1980s emerges the idea that a global overview of the protein's energy surface is of paramount importance 
for a quantitative understanding of the relationships between structure, dynamics, stability, and functional behavior 
of proteins \cite{Ansari-PNAS1985, Frauenfelder-ARBC1988, Frauenfelder-Science1991, Onuchic-AnnuRevPhysChem1997, 
Kitao-Proteins1998, Brooks-Science2001, Krivov-PNAS2004, Wales-JPCB2006, Senet-PNAS2008}. Thanks to the continuous 
increase of the computing power\cite{Shaw-JACS2012, Piana-PNAS2012} and of the reliability of empirical 
force fields \cite{Adcock-ChemRev2006, Beauchamp-JCTC2012}, all-atom molecular dynamics (MD) simulations become a widely 
employed computational technique to simulate the dynamics of complex systems such as proteins through discrete 
integration of the Newton's equations of motions of each atom. Under the assomption of ergodicity one can then derive 
the equilibrium properties of interest of a protein as time averages of the corresponding observables along the MD trajectories 
and compute the associated free-energy. However in several cases all-atom MD simulations are still not competitive 
to describe the protein conformational dynamics, due to the fact that using an atomistic model is computationally expensive, 
as sufficiently realistic potential energy functions are intrinsically complex. Moreover, most phenomena of interest take place 
on time scales that are orders of magnitude larger than the accessible time that can be currently simulated with classical 
all-atom molecular dynamics \cite{Bowman-JACS2011}.
This issue can be addressed either by reducing the dimension of the conformational space explored by the protein during 
the MD simulations by using a coarse-grained representation of the protein structure, where each residue has only a few degrees 
of freedom and where the solvent surrounding the protein is implicit \cite{Liwo-JPCB2007, Kortuta-PNAS2009}, 
or by accelerating the exploration of the conformational space in (all-atom) MD simulations. In the latter case, 
a large variety of methods referred to as \textit{enhanced sampling techniques} have been proposed 
\cite{Carter-CPL1989, Bash-Science1987, Patey-JCP1975, Grubmuller-PRE1995, Ferrenberg-PRL1989, Jarzynski-PRL1997, 
Darve-JCP2001, Gullingsrud-JCP1999, Huber-JCAMD1994, Rosso-JCP2002}. They exploit a methodology aimed at accelerating 
rare events and based on constrained MD. 
Metadynamics (metaD) \cite{Laio-PNAS2002, Laio-RPP2008, Barducci-Wiley2011} belongs to this class of methods: 
it enhances the sampling of the conformational space of a system along a few selected degrees of freedom, 
named collective variables (CVs) and reconstructs the probability distribution as a function of these CVs. 
In a nutshell, the dynamics in the space of the chosen CVs is enhanced by disfavoring already visited regions through the use 
of a history-dependent potential, constructed as a sum of Gaussians centered along the trajectory followed by the CVs. 
This sum of Gaussians is then exploited to reconstruct iteratively an estimator of the free-energy surface spanned by the CVs 
in the region explored during the biased dynamics. 
Well-Tempered Metadynamics \cite{Barducci-PRL2008} (WTmetaD) is the most recent variant of the method, which, 
because of its convergence properties, is the most widely adopted version of the metadynamics algorithm. In WTmetaD, the bias 
deposition rate decreases over simulation time and the dynamics of all the microscopic variables becomes progressively 
closer to thermodynamic equilibrium as the simulation proceeds, making the bias to converge to its limiting value 
in a single run and avoiding the problem of overfilling, \textit{i.e.} when the height of the accumulated Gaussians largely exceeds 
the true barrier height.

However, WTmetaD suffers from the same limitation than standard metaD, \textit{i.e.} its succes depends on the critical choice 
of a reasonable number of relevant CVs. All the relevant slow varying degrees of freedom must be catched by the CVs. 
In addition, the number of CVs must be small enough to avoid exceedingly long computational time, while being able to distinguish 
among the different conformational states of the system. Consequently, identifying 
a set of CVs appropriate for describing complex processes involves a right understanding of the physics 
and chemistry of the process under study \cite{Branduardi-JACS2005}. Choosing a correct set of CVs thus 
remains a challenge, as a whole, independently of the enhanced sampling technique one could consider.
A way to overcome the difficulties to extract the functionally relevant motions from MD simulations is to use 
collective coordinates identifying a low dimensional subspace in which the significant, functional 
protein motions are expected to take place. Principal Component Analysis (PCA) of the structure fluctuations 
of a protein, also called \textit{essential dynamics}, is a powerfull method to represent 
in a space of only a few collective coordinates (typically between 1 and 3) both the functional modes of proteins 
in their native state \cite{Ichiye-Proteins1991, Garcia-PRL1992, Amadei-Proteins1993, Kitao-COSB1999, 
Groot-JMB2001, Hall-COCB2008, Daidone-Wiley2012} and the folding pathways of certain 
proteins \cite{Maisuradze-Proteins2007, Maisuradze-JMB2009}. Therefore PCA emerges in the last few years 
as a relevant candidate to reconstruct the protein free-energy landscape both with standard metaD 
and WTmetaD \cite{Spiwok-JPCB2007, Spiwok-JMM2008, Sutto-JCTC2010}. 
These methods show the efficience of this particular class of CVs in terms of convergence time 
and free-energy reconstruction accuracy. However it is well known that the smooth appearance of the FES obtained with 
PCA of the fluctuations of the Cartesian coordinates of the atoms often represents an artifact of the mixing of the internal 
and overall motions of the protein \cite{Mu-Prot2005}. Indeed, whereas the overall translation of a molecule can readily be separated 
from its internal deformations by requiring that the molecular center of mass is fixed, the elimination of overall rotation 
of the molecule from its internal motions is only possible for relatively rigid molecular structure, 
\textit{i.e.} for which the structural fluctuations can be described as small departures from a reference molecular structure. 
In this case, the separation can be achieved by superimposing the structures at the snapshots of a MD trajectory 
to a reference structure, such that the root-mean-square deviations of the fluctuations of the atom positions 
is minimized \cite{Diamond-PS1992}. While this procedure correctly removes the overall rotation 
in the limit of small fluctuations, it breaks down in the case of large amplitude motions, where it is not possible 
to unambiguously define a single reference structure. 
It is then interesting to consider the use of linear correlation analysis for internal coordinates, for which the overall motion 
is safely eliminated without the use of any arbitrary reference structure. The dihedral angles are appropriate internal 
coordinates for biopolymers. However, attention has to be paid to the periodicity of 
such quantities as the arithmetic mean of dihedral angles can not be easily calculated as in Cartesian coordinates. 
One can circumvent this problem by considering the method referred to as \textit{dihedral Principal Component analysis} (dPCA) 
based on (Ramachadran) dihedral angles \cite{Mu-Prot2005, Altis-JCP2007}. In dPCA, one must perform a transformation 
from the space of dihedral angles to a metric coordinate space built up by the trigonometric functions of the dihedral angles. 
The resulting representation of the dihedral angles is unique and therefore allows us to readily calculate the mean 
and the covariance matrix.

A constraint of using PCA (or dPCA) to select the relevant CVs representing the functional motions of a protein, is that 
some prior knowledge of the dynamics of the system under study is necessary. 
The structural fluctuations of the molecule must be determined from a preliminary short unbiased MD simulation to 
which dPCA is applied.

In the present paper we perform biased ($250$ ns) and unbiased ($2.6 \,\mu s$) simulations of a penta-peptide, 
Met-enkephalin (Tyr-Gly-Gly-Phe-Met), which is one of the smallest neurotransmitter peptides \cite{Hughes-Nature1975}. 
Because of its important roles in physiological processes as for examples the mediation of pain and respiratory 
depression, this neuropeptide has been the subject of numerous studies both experimental 
\cite{Khaled-BBRC1976, Spirtes-BBRC1978, Smith-Science1978, Schiller-BBRC1983, Rapaka-ABL1985, Graham-Biopoly1992, DAlagni-EJB1996} and numerical \cite{Sanbonmatsu-Prot2002, Shen-BJ2002, Henin-JCTC2010, 
Sutto-JCTC2010, Chen-JCP2012}. This small flexible peptide does not adopt a single conformation in aqueous solution, 
that is consistent with the requirement that Met-enkephalin be sufficiently flexible to bind to different opioid 
receptors \cite{Graham-Biopoly1992}, and its free-energy surface is complex and well-structured. 
However many aspects of both the structure and dynamics of Met-enkephalin in aqueous solution remain unsolved. 

We show that coupling WTmetaD with the dPCA of the Ramachadran dihedral angles 
provides a rugged FEL of Met-enkephalin, which is very different from 
the result one can extract from the coupling of WTmetaD with the PCA of the Cartesian coordinates of the atoms. 
We demonstrate that combining WTmetaD with dPCA speeds up the reconstruction of a free-energy surface 
of the peptide compared with the unbiased MD simulations.  
Finally, we underline a criterion to limit the number of CVs necessary to biased the metadynamics simulation 
in terms of the existence of the minima of the one-dimensional free-energy profiles associated to 
Ramachandran dihedral angles along the amino-acid sequence.

\section{Computational Method}
\textbf{MD simulation.} All-atom MD simulations in explicit water of Met-enkephalin 
(PDB ID: 1PLW) were carried out with the version 4.5.5 of the GROMACS software package \cite{Hess-JCTC2008} 
using TIP3P water model \cite{Jorgensen-JCP1983} and the Amber99SB force field \cite{Hornak-Proteins2006}. 
Biased simulations were performed using version 1.3 of the plugin for free-energy calculation, named PLUMED 
\cite{Bonomi-CPC2009}. The Met-enkephalin was solvated in 892 TIP3P water molecules enclosed in a cubic box of 27.7 $nm^3$ 
under periodic boundary conditions. The time step used in all simulations was 0.001 ps, and the list of neighbors 
was updated every 0.005 ps with the ``grid'' method and a cutoff radius of 1.4 nm. 
The coordinates of all the atoms in the simulation box were saved every 1 ps. The initial velocities were chosen randomly. 
We used the NPT ensemble with the temperature and pressure kept to the desired value by using the Beredsen method 
and an isotropic coupling for the pressure ($T=300$K, $\tau_T = 1$ fs; $P_O = 1$ bar, coupling time $\tau_P=1$ ps). 
The electrostatic interactions were computed by using the particle mesh Ewald (PME) algorithm 
\cite{Darden-JCP1993, Essman-JCP1995, Kawata-CPL2001} (with a radius of 1 nm) 
with the fast Fourier transform optimization. The cutoff algorithm was applied for the noncoulomb potentials 
with a radius of 1 nm.  The energy of the model was first optimized with the ``steepest descent minimization'' algorithm 
and then by using the ``conjugate gradients'' algorithm. The system was warmed up for 40 ps and equilibrated for 200 ps 
with lower restraints, finishing with equilibration without restraints at 300 K for 1 ns. 
The simulation was then continued for 2.6 $\mu s$.

\textbf{Determination of metadynamics CVs.} The set of CVs used in WTmetaD simulations 
were automatically generated from a dPCA of the 8 backbone Ramachadran dihedral angles \cite{Ramachandran-JMB1963} 
of the unbiased MD trajectory. In short, dPCA consists in diagonalizing the covariance matrix 
of the Cartesian components of the two-dimensional vectors $\vec{q}(t)$ representing the set of dihedral angles 
computed from a preliminary unbiased MD run, say $\vec{q}(t) \equiv \{q_{4n} = \cos\phi_n, 
\, q_{4n-1} = \sin\phi_n, \, q_{4n-2} = \cos\psi_n,\, q_{4n-3} = \sin\psi_n\}$ with $1 \leq n \leq 4$ 
in the case of Met-enkephalin. The eigenvalues $\lambda_k$ of the covariance matrix 
are ordered by decreasing value and each collective mode $k$ is characterized by $\lambda_k$ and by the corresponding eigenvector $\vec{e}^{(k)}$. 
The eigenvectors corresponding to the largest eigenvalues contain the largest fluctuations and hence hold the most 
important variability of the system. The contribution of the backbone dihedral angles $\phi_n$ and $\psi_n$ to a mode $k$ 
is quantified by the so-called influence \cite{Altis-JCP2007} $\nu_n^{(k)}$: 
\begin{equation}\label{InfluenceDefinition}
 \nu_n^{(k)} \equiv \left\{
  \begin{array}{ll}
     \Big(e^{(k)}_{4n}\Big)^2 + \Big(e^{(k)}_{4n-1}\Big)^2 & \mbox{for }  \phi_n \\
     {} \\
     \Big(e^{(k)}_{4n-2}\Big)^2 + \Big(e^{(k)}_{4n-3}\Big)^2  & \mbox{for }  \psi_n.
  \end{array}
 \right.
\end{equation}
The largest values of the influence $\nu_n^{(k)}$ reveal the dihedral angles which contribute the most to the fluctuations in this mode. 
One or more principal components $PC^{(k)}$ associated to eigenvectors $\vec{e}^{(k)}$, \textit{i.e.} the projection 
of the trajectory $\vec{q}(t)$ on the respective eigenvectors:
\begin{equation} \label{PrincipalComponents}
 PC^{(k)}(t) = \sum_{n=1..16} \Big( q_n(t)-\langle q_n \rangle \Big)\times e^{(k)}_{n},
\end{equation}
where $\langle \dots \rangle$ denotes the average over the duration of the trajectory, 
were selected as CVs in WTmetaD trajectories of the pentapeptide. Considering the 2.6 $\mu s$ unbiased MD simulation 
the four largest eigenvalues of the covariance matrix ($\lambda_1, \lambda_2, \lambda_3$ and $\lambda_4$) describe 
$18\%$, $16\%$, $13\%$ and $11\%$ of the fluctuations with cosine contents \cite{Hess-PRE2000} 
of $0.01$, $0.006$, $0.004$ and $0.003$ respectively. 
The same dPCA repeated using only the first 26 ns or 260 ns ($1\%$ and $10\%$ of the simulation respectively) of the 
unbiased trajectory gave similar results. In particular, the overlap between the subspace generated by the first two 
eigenvectors ($\vec{e}^{(1)}$, $\vec{e}^{(2)}$) calculated after the first 26 ns or 260 ns of the unbiased MD trajectory 
and the reference subspace generated by the first two eigenvectors of the 
whole 2.6 $\mu s$ trajectory are $0.77$ and $0.96$ respectively. 
The overlap is calculated as the root-mean-square inner product of the first two PCA eigenvectors \cite{Amadei-Prot1999}. 
Each $PC^{(k)}$ is associated with a free-energy profile (FEP) $V_k$ corresponding to the projection of the FEL of
the protein along the collective coordinate $PC^{(k)}$, and computed by using the usual Boltzmann formula
\begin{equation} \label{BoltzmannFormula1}
 V_k = -kT \, \ln\Big[P\Big(PC^{(k)}\Big)\Big]
\end{equation}
where $P(PC^{(k)})$ is the probability density of the variable $PC^{(k)}$ computed over the trajectory 
considered, $k$ is the Boltzmann constant and $T$ is the temperature. Similarly, we calculated the 
free-energy surface (FES) based on the two first principal components as 
\begin{equation} \label{BoltzmannFormula2}
 V_{1,2} = -kT \, \ln\Big[P\Big(PC^{(1)}, PC^{(2)}\Big)\Big]
\end{equation}
where $P(PC^{(1)}, PC^{(2)})$ is the two-dimensional probability density computed from the trajectory considered.

\textbf{WTmetaD simulation.} The well-tempered variant of the metadynamics enhanced sampling technique 
was used for the all-atom biased MD simulations of Met-enkephalin using the CVs selected from the dPCA 
of the unbiased MD trajectory. According to the algorithm introduced by Barducci et al. \cite{Barducci-PRL2008} 
a Gaussian is deposited every $\tau_G = 4 ps$ with height $W = W_0 e^{-V(s,t)/(f-1)T}$, where $W_0 = 2$ kJ/mol 
is the initial height, $T$ is the temperature of the simulation and $f \equiv (T+\Delta T)/T = 1.5$ the bias factor 
with $\Delta T$ a parameter with the dimension of a temperature. The resolution of the recovered free-energy surface 
is determined by the width of the Gaussians $\sigma = 0.1$ in units of the respective CV. 
We performed 2 runs of WTmetaD of Met-enkephalin, each of $250$ ns, using the dPCA eigenvectors 
generated from both the whole $2.6 \, \mu s$ MD trajectory and the shorter trajectory of $26$ ns ($1\%$ of the 
whole unbiased MD simulation).

\textbf{Comparison of potential energy functions.} To quantitatively compare the result of the metadynamics simulation to the reference unbiased molecular dynamics, 
we consider the distance measure introduced by Alonso and Echenique \cite{Alonso-JCC2006} and previously used 
in Sutto et al. \cite{Sutto-JCTC2010} to compare two different energy functions, 
\textit{i.e.} $V_{met}(s)$ and $V_{ref}(s)$ expressed in terms of the CVs defined in a region $\Omega$: 
\begin{equation}
 d_A(V_{met},V_{ref}) = \Big( (\sigma_{met}^2 + \sigma_{ref}^2) (1-r_{met,ref}^2) \Big)^{1/2} ,
\end{equation}
where $\sigma_x$, with $x$ denoting either $met$ or $ref$, is the statistical variance of the free energy $V_x$ 
defined by $\sigma_x^2 = \frac{1}{N}\int_{\Omega}ds \, (V_x(s)-\langle V_x \rangle)^2$. 
$\langle V_x \rangle = \frac{1}{N}\int_{\Omega}ds \, V_x(s)$ is the average value of $V_x$ in the region $\Omega$ 
and $N=\int_{\Omega} ds$ is a normalization constant. The variance $\sigma_x$ sets the physical scale of the measure 
and confers the energy units to the distance. $r_{met,ref}$ is the Pearson correlation coefficient and measures 
the degree of correlation between the two energy surfaces. It is defined by  $r_{met,ref}=cov(V_{met},V_{ref})/\sigma_{met}\sigma_{ref}$
where $cov(V_{met},V_{ref})$ is the covariance between the two energies. The $d_A$ measure is convenient 
since it is expressed in energy units and can be directly compared to the thermal fluctuations. 

\textbf{Clustering analysis.} To compute the representative structures of Met-enkephalin 
in a cluster, \textit{i.e.} in a local minima in a predefined space, a \textit{distance measure} between structures 
is defined considering $C^{\alpha}$-RMSD after fitting to an arbitrary structure belonging to the local minimum. 
The following procedure is used \cite{Nicolai-JBSD2012}. First, the RMSD ($C^{\alpha}$-RMSD) between 
the coordinates of the $C^{\alpha}$ atoms of an arbitrary chosen structure in the cluster and the coordinates 
of the $C^{\alpha}$ atoms of all the other structures in the cluster is computed. Second, the one-dimensional 
probability distribution $P(C^{\alpha}-RMSD)$ associated to the cluster is calculated and the coordinates of 
Met-enkephalin of all snapshots corresponding to the maximum value of $P(C^{\alpha}-RMSD)$ are extracted from 
the trajectory. This leads to a set of representative structures (most probable) for the MD run considered 
associated to the cluster. Third, the average structure of this cluster is computed and the RMSD between 
this average structure and all the members of the set is calculated. The structure within the set of 
representative structures having the minimum RMSD relative to the average structure of the set is chosen 
as the representative structure of the cluster.

\section{Results}
\begin{figure}
 \includegraphics[width=1.0 \columnwidth, angle=-0]{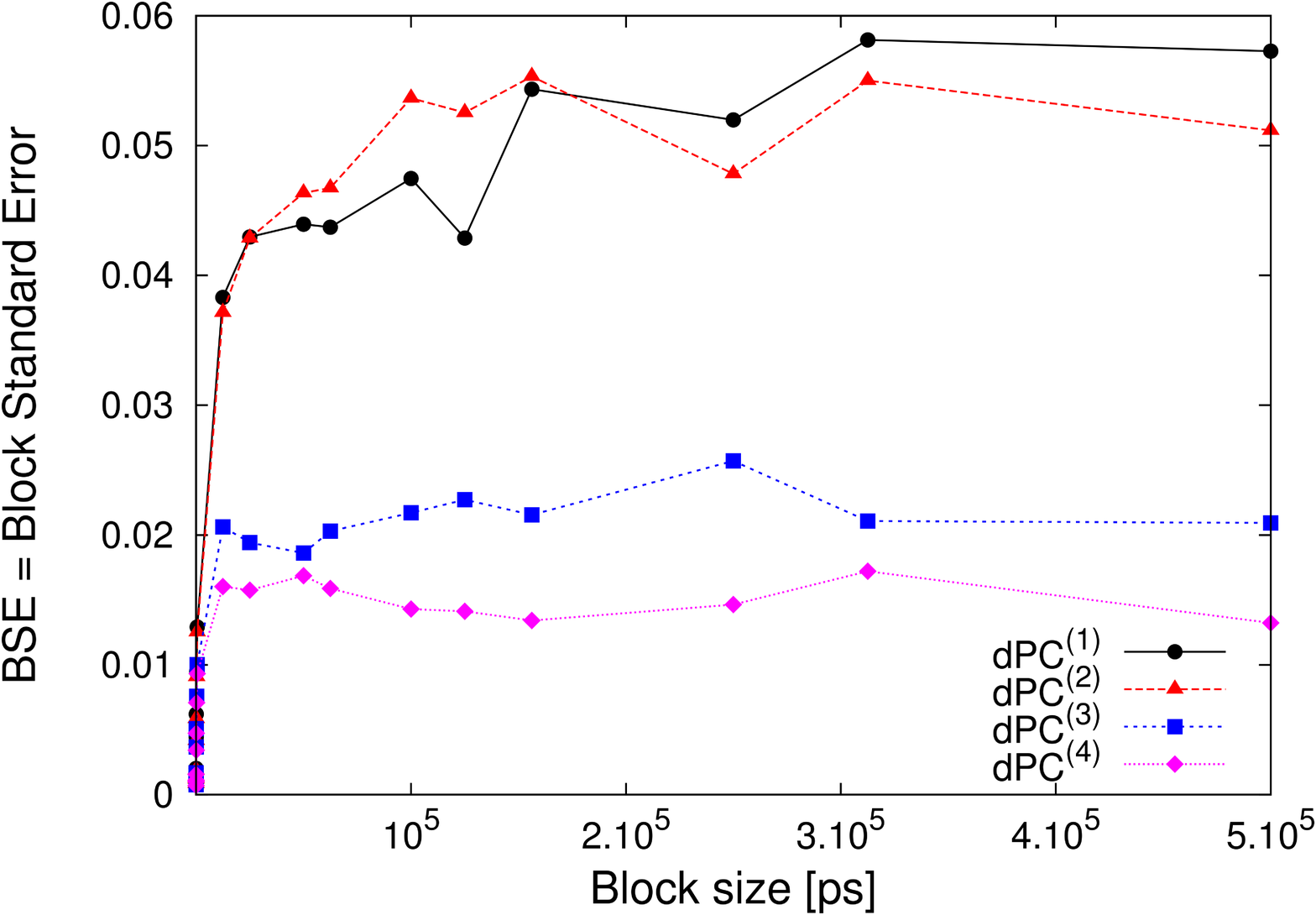}
 \caption{Block Standard Error associated with the first four dihedral principal components 
 for the reference $2.6 \,\mu s$ unbiased MD trajectory of Met-enkephalin. The functions increase 
 monotonically with the block size and converge asymptotically to a constant which is the true 
 standard error associated with $\langle dPC^{(i)} \rangle$.}
 \label{BSE_dPCA}
\end{figure}

\textbf{Reference free-energy surface.} The reference FES is obtained from a 2.6 $\mu s$ unbiased simulation, 
whose convergence with respect to the trigonometric functions of the 8 backbone Ramachandran dihedral angles of 
Met-enkephalin was checked by applying the block analysis \cite{Grossfield-ARCC2009} to the first dihedral principal 
components, namely dPC$^{(i)}(t)$ with $i$ from $1$ to $4$ (cf. \ref{PrincipalComponents}). 
This procedure consists in dividing the trajectory into $M$ time segments (or blocks) and in considering a full range 
of block sizes. Small blocks will tend to be highly correlated with neighboring blocks, whereas blocks longer than 
important correlation times will only be weakly correlated. The true statistical uncertainty, \textit{i.e.} 
the standar error, is obtained when the Block Standard Error (BSE), defined as $BSE(n) = \frac{\sigma_n}{\sqrt{M}}$, 
ceases to vary and reaches a plateau, reflecting essentialy independent blocks. $\sigma_n$ represents the standard 
deviation among the block averages and $n$ is the size of the blocks (the number of frames in each block). 
In \ref{BSE_dPCA} are shown the BSEs associated with the first four dihedral principal components for 
the reference 2.6 $\mu s$ unbiased trajectory. We see that the first dPC (dPC$^{(1)}$) is the one which presents 
a BSE with the slowest convergence, reaching the plateau characterizing the associated true standard error approximately 
for a block size of $200$-$300$ ns. In comparison, the BSEs associated with dPC$^{(2)}$ to dPC$^{(4)}$ reach a plateau 
for block size of $150$ ns and $10$ ns respectively. Let us note that this is coherent with the property of PCA 
for which the first PC captures the slowest mode corresponding to the maximum of variability.

\begin{figure}
 \includegraphics[width=1 \columnwidth, angle=0]{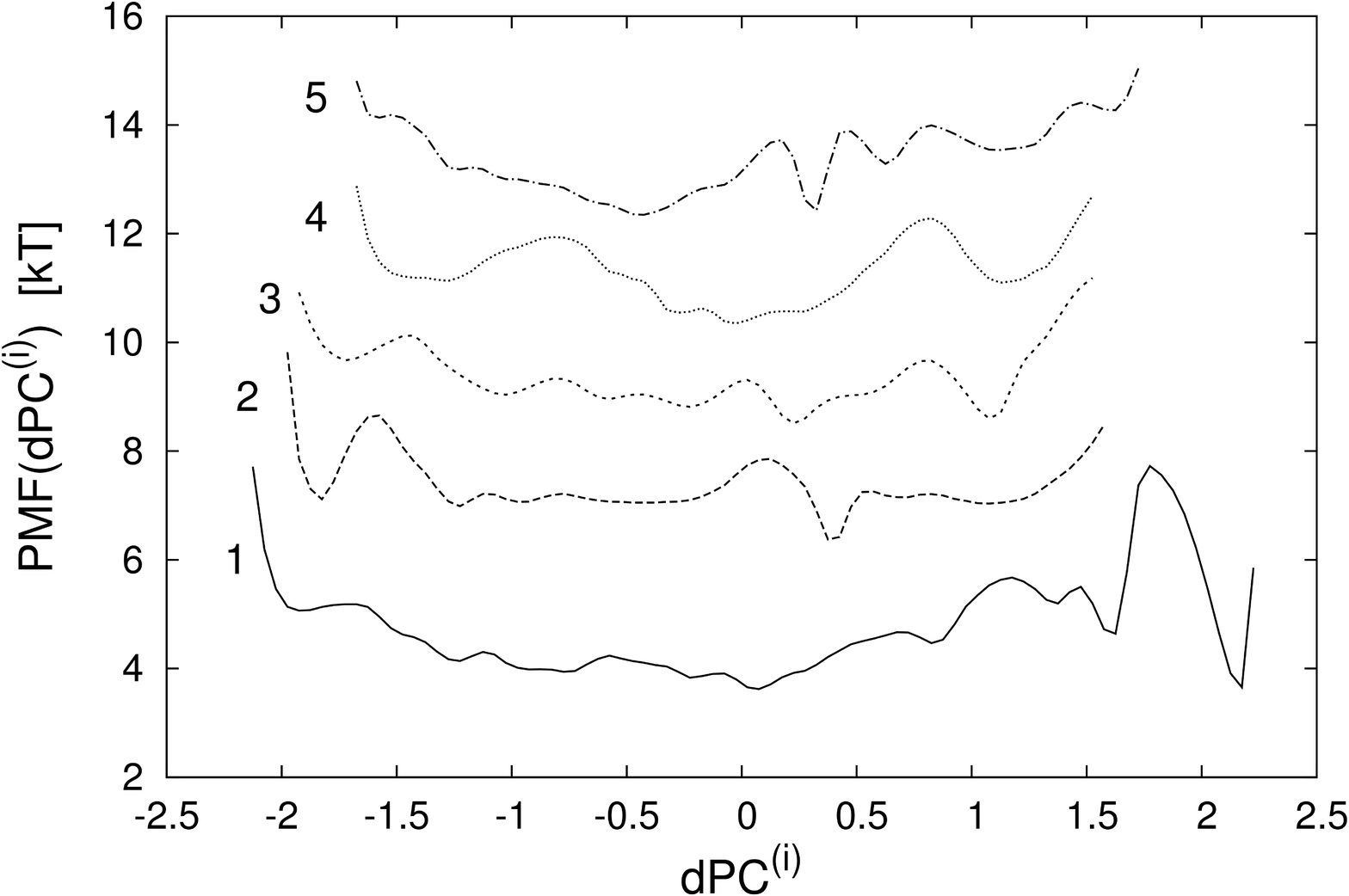}
 \caption{Free-energy profiles of the first five dihedral principal components for the reference 
 $2.6 \, \mu s$ unbiased trajectory. The numbers 1 to 5 in the panel refer to label of the dPCs. 
 The free-energy profiles are shifted arbitrarily along the ordinate axis for clarity.}
 \label{hierarchicaldPCA}
\end{figure}

\begin{figure*}
 \includegraphics[width=2.1\columnwidth, angle=0]{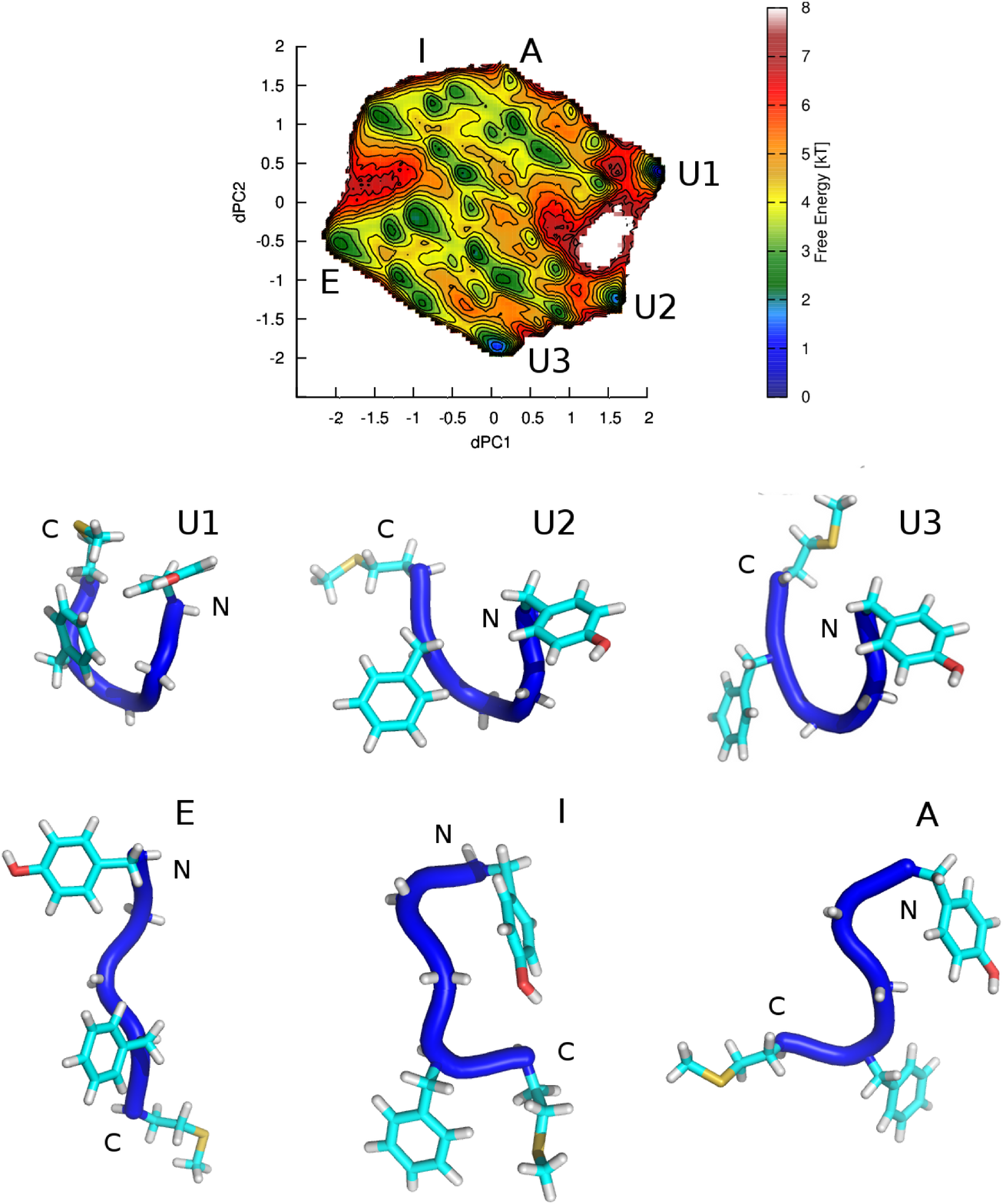}
 \caption{Clutering analysis of the free-energy surface (cf. \ref{BoltzmannFormula2}) as a function of 
 the projection over the first two principal eigenvectors computed from the whole unbiased MD trajectory 
 of Met-enkephalin. The contour lines are every $0.5 \,k_BT$. The representative conformations (most probable) correspond 
 to U-shaped backbone with and without packed aromatic rings ($U1$, $U2$ and $U3$), elongated conformation ($E$), 
 helix-like structure ($A$) and an intermediate structure ($I$). N and C indicate the N-term and C-term of 
 the peptide, respectively. This figure has been prepared using Pymol (http://www.pymol.org)}
 \label{clustering}
\end{figure*}

The principal component analysis also gives information about the dimension of the space in which the conformational 
substates of the protein are distributed. In this spirit, Kitao et al. \cite{Kitao-Proteins1998} proposed a classification 
of  principal modes into three types of modes: multiply-hierarchical, singly-hierarchical and harmonic modes. 
This classification is based on the properties of the FEPs $V_k$ computed from the whole unbiased MD trajectory 
using the Boltzmann formula (cf. \ref{BoltzmannFormula1}). Multiply-hierarchical, singly-hierarchical and harmonic modes 
have a FEP with multiple basins of minima, a FEP with a single basin of minima, or a single minima harmonic FEP, respectively.
Considering the approach of Hegger et al. \cite{Hegger-PRL2007} the dimension of the FEL  
is then defined by the number of multiply-hierarchical PCs. \ref{hierarchicaldPCA} illustrates the FEPs 
of the first five dPCs characterized as multiply-hierarchical (\textit{i.e.} they contain more than 
one major basin of minima). For more clarity, the FEPs are shifted arbitrarily along the ordinate axis. 
This means that dPC$^{(i)}$ with $i$ from $1$ to $5$ are the main contributors to the structural fluctuations of the pentapeptide, 
and are associated with global collective motions of the system, \textit{i.e.} they contain the most large conformational fluctuations. 
Indeed the protein moving along a multiply-hierarchical principal component significantly changes its intra-molecular packing topology. 
The other dPCs, namely dPC$^{(6)}$ to dPC$^{(16)}$, have either singly hierarchical FEPs mainly related to local 
fluctuations of side chains, or harmonic FEPs involving local motions of the backbone that do not contribute significantly to global 
conformational changes of Met-enkephalin.
It emerges from the analysis of \ref{hierarchicaldPCA} that information about the first five dPCs is necessary 
to discriminate uniquely the conformational states of the protein. 
It is nevertheless instructive to analyse the two-dimensional free-energy map (dPC$^{(1)}$, dPC$^{(2)}$) to link 
the present analysis to the one obtained with a Cartesian PCA in Sutto et al.\cite{Sutto-JCTC2010}, 
and to compare our sampled conformations to the litterature values. Visual inspection of \ref{clustering} 
mainly exhibits a quite rugged and complex free-energy landscape which contains numerous minima. 
This result is different from the smooth appearance and the simple single basin of minima with a funnel-like shape observed 
in Sutto et al.\cite{Sutto-JCTC2010} where the Cartesian PCA only reveals 3 shallow minima. 
This is however not surprising as the same kind of observations have already been stated by Mu et al.\cite{Mu-Prot2005} 
in the case of penta-alanine who related this difference to the artifact of the unavoiding mixing of the internal and overall 
motion of the peptide in the Cartesian PCA of its trajectory.

Clustering analysis of the (dPC$^{(1)}$, dPC$^{(2)}$) FES (see Computational method) represented in \ref{clustering} 
highlights three main basins ($U1$, $U2$ and $U3$) related to extreme values of dPC$^{(1)}$ and dPC$^{(2)}$ and isolated from the rest 
of the FES with relevant barrier, and correponding to a set of U-shaped conformations found by Sanbonmatsu and Garcia 
\cite{Sanbonmatsu-Prot2002} and by Henin et al.\cite{Henin-JCTC2010}. Let us note that this is different 
from the Cartesian PCA of Sutto et al.\cite{Sutto-JCTC2010}, who found a single basin in their FES calculation 
corresponding to the U-shaped conformations of Met-enkephalin. 
As shown in \ref{clustering}, the basin $U1$ of the (dPC$^{(1)}$, dPC$^{(2)}$) FES corresponds to a U-shaped 
conformation in which the phenylalanine and tyrosine side chains are packed. A move from the $U1$ basin 
to the $U2$ and $U3$ basins corresponds to torsional deformations of the peptide leading to the unpack 
of the phenylalanine and tyrosine side chains. In the torsional deformation from $U1$ to $U3$, 
the left and right arms of the U-shaped conformations become aligned causing 
the midsection of the backbone to untwist. 
The passage from one U-shape basin to another requires crossing the largest activation barriers 
(of about $3\,kT = 7.5$ kJ/mol) on the FES shown in \ref{clustering}. Except the basins U1, U2 and U3, 
the rest of the FES contains a numerous minima of comparable probabilities and lifetime, in particular more elongated 
structures ($E$) and helix-like conformations ($A$) also in agreement with the aforementionned work\cite{Sutto-JCTC2010}. 
\begin{figure*}
 \includegraphics[width=0.85\columnwidth, angle=0]{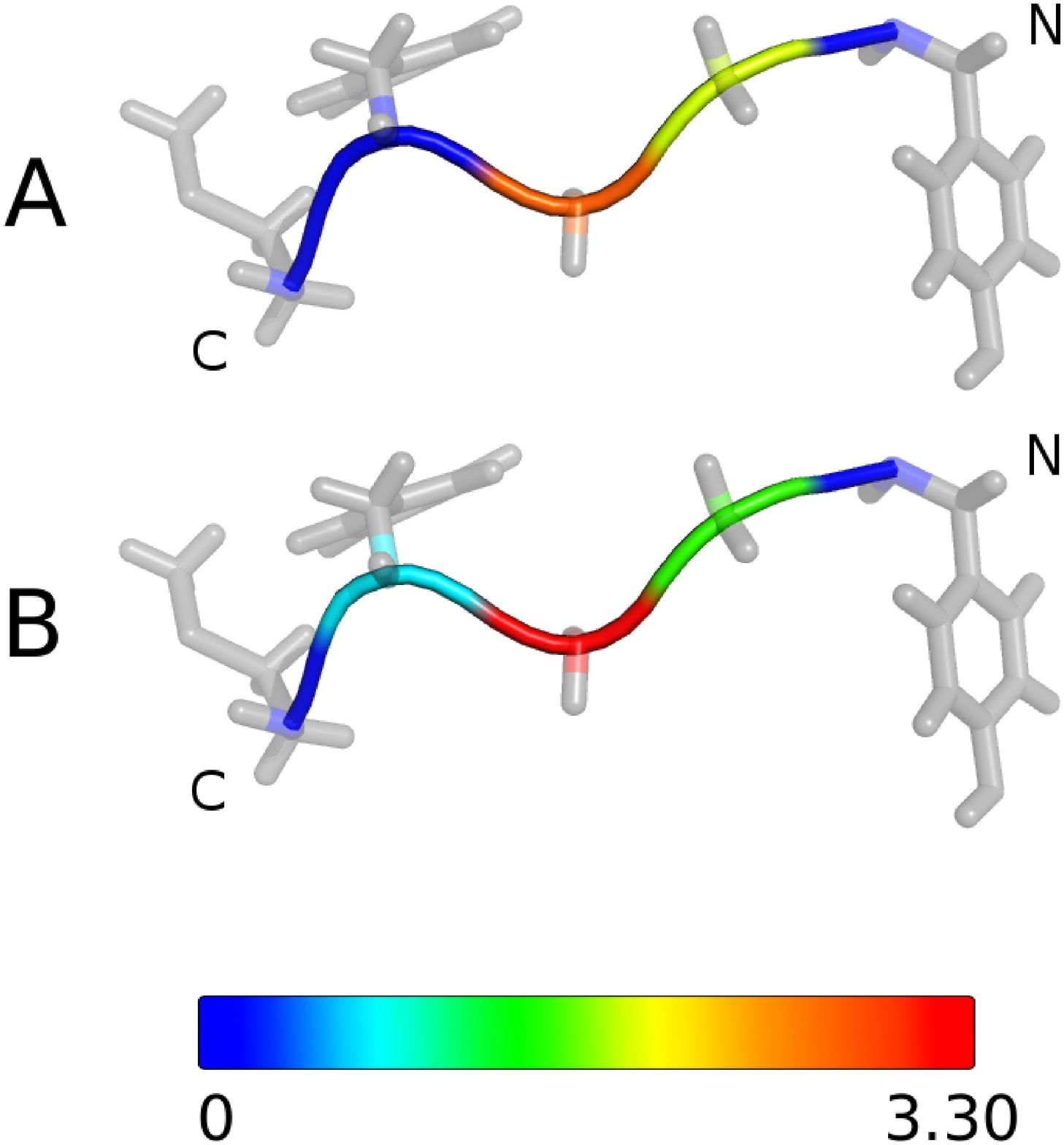}
 \hskip 1.0cm
 \includegraphics[width=0.85\columnwidth, angle=0]{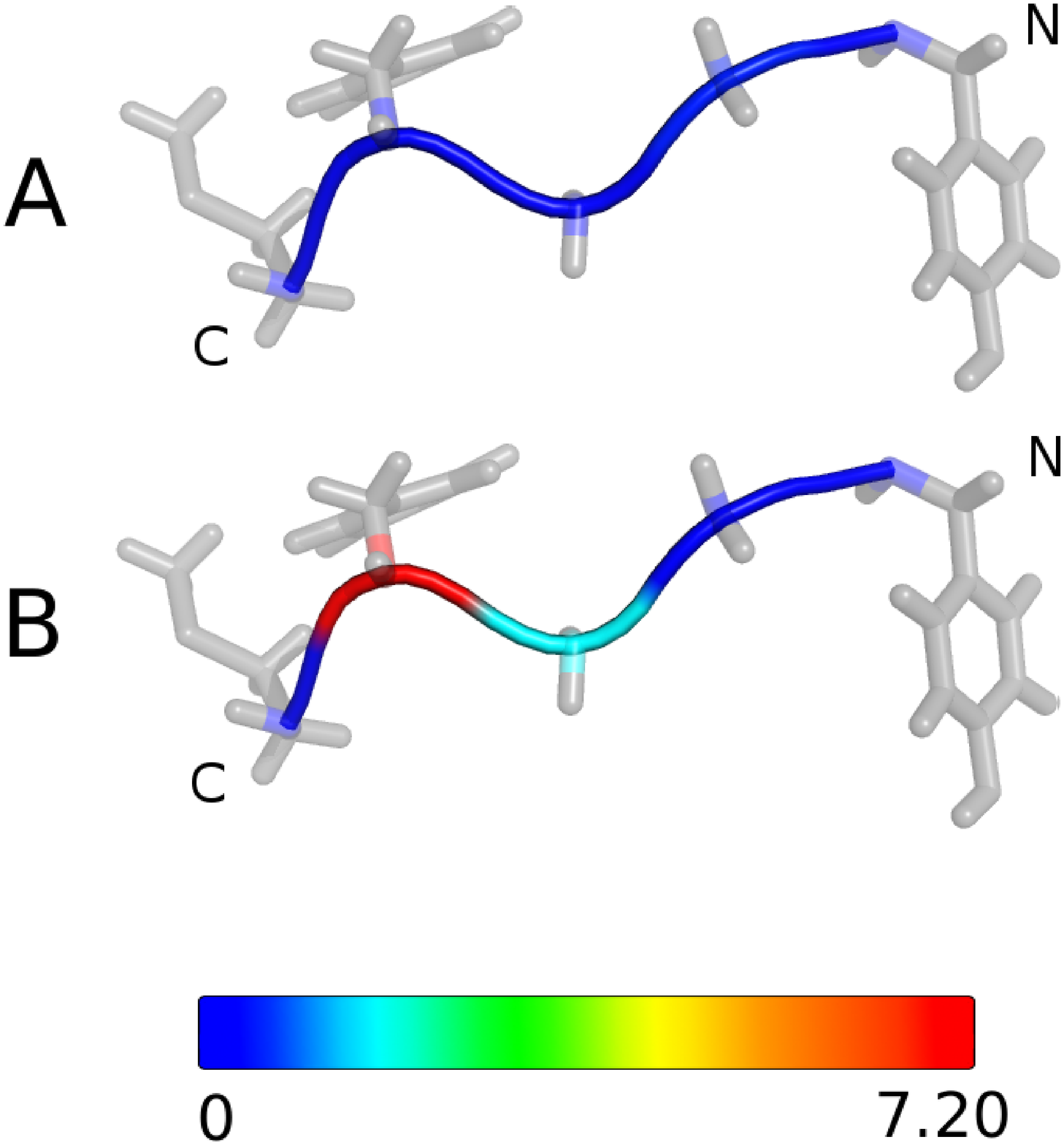}
 \caption{Influence of the first two dihedral-PC, dPC$^{(1)}$ (left panel) and dPC$^{(2)}$ (right panel) defined as the contribution 
 of the backbone dihedral angles $\Phi_n$ (panels A) ans $\Psi_n$ (panel B) to a mode $k$. The largest values of the influence 
 (red color) reveal the dihedral angles which contribute the most to the fluctuations in this mode. 
 N and C indicate the N-term and C-term of the peptide, respectively. This figure has been prepared using Pymol 
 (http://www.pymol.org)}
 \label{influ_dPC1dPC2}
\end{figure*}
The relation between the dPC$^{(1)}$ and dPC$^{(2)}$ and the Ramachandran dihedral angles along the sequence of 
the peptide is given by the value of their influence (cf. \ref{InfluenceDefinition})  
in the eigenvector $\vec{e}^{(1)}$ and $\vec{e}^{(2)}$, respectively. 
In the mode 1, the influence is the largest for the couple of Ramachandran dihedral angles 
of GLY3 (highest) and GLY2 (cf. \ref{influ_dPC1dPC2}).  Not surpringly, the small size of 
the H side-chain of the GLY residues permit at the middle part of the peptide chain to be very flexible. 
In mode 2, the highest influence is for the Ramachandran dihedral angle $\Psi$ of PHE4, followed by 
the influence of $\Psi$ of GLY3. The system then undergoes simultaneous torsional and compressional 
deformations along the axis defined as the two first dPCA eigenvectors and illustrated through the related 
influence in \ref{influ_dPC1dPC2}.

\begin{figure*}
 \includegraphics[width=0.8\textwidth, angle=0]{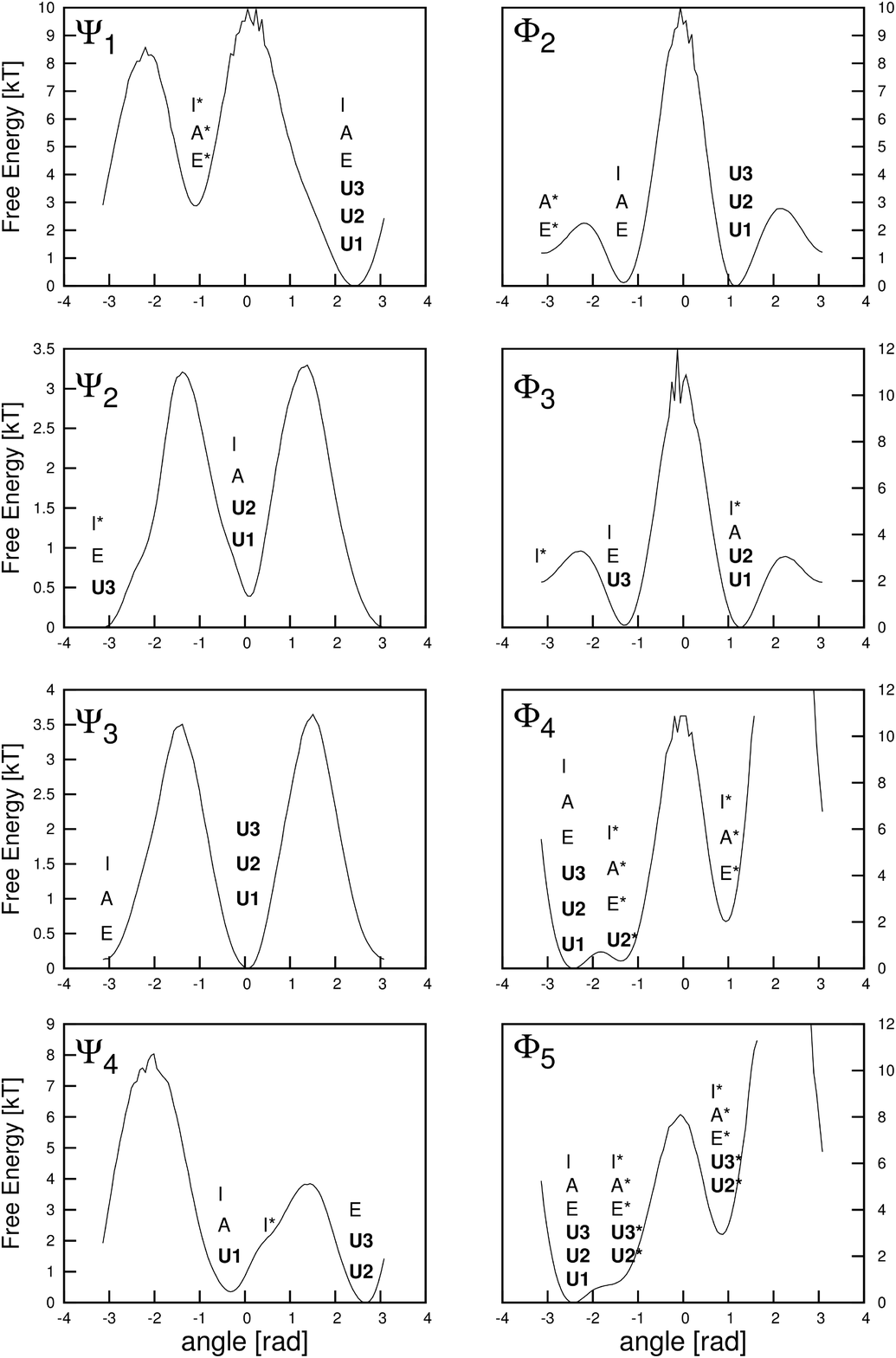}
 \caption{Representaion of the eight one-dimensional free-energy profiles of the Ramachandran dihedral angles 
 for the reference $2.6 \, \mu s$ unbiased simulation. $U1$, $U2$, $U3$, $A$, $E$, and $I$ denote the values of the Ramachandran 
 dihedral angles of the corresponding representative (most probable) structures shown in \ref{clustering}. 
 $U1^*$, $U2^*$, $U3^*$, $A^*$, $E^*$ and $I^*$ 
 denote other structures in the basins $U1$, $U2$, $U3$, $A$, $E$ and $I$ within $0.5\,k_B$T from the global minima 
 of the corresponding basins in \ref{clustering}.}
 \label{FEP_PhiPsi}
\end{figure*}

To enforce the pertinence of the analysis in the two-dimensional essential space to explore the 
conformational space, we link the clustering analysis shown in \ref{clustering} with the study of the one-dimensional 
FEPs of the 8 Ramachandran dihedral angles represented in \ref{FEP_PhiPsi}. 
A similar approach was recently used to study the dynamics of side-chains of a protein in its native 
state \cite{Cote-PNAS2012}. Indeed, by mapping the different points of the two-dimensional essential 
free-energy surface on the set of the free-energy profiles along the amino-acid sequence of the peptide,
we obtain a representation of the configurational changes in a complete basis, \textit{i.e.} 
allowing the reconstruction of the backbone of the pentapeptide. It is relevant to notice that the
representative (most probable) structures associated to the three main basins 
$U1$, $U2$ and $U3$ occupy the global minima in the one-dimensional FEPs of the $8$ Ramachandran dihedral 
angles or metastable states very close to these global minima as for $U1$, $U2$ for $\Psi_2$ and for $U1$ for $\Psi_4$
(cf. \ref{FEP_PhiPsi}). 
Moreover, if we consider the set of 
structures belonging to these three basins $U1$, $U2$ and $U3$ (cf. \ref{clustering}), we see that other local minima are filled 
(denoted $U1^*$, $U2^*$ and $U3^*$ in \ref{FEP_PhiPsi}) . Indeed, if we do not restrict the present analysis 
to the basins $U1$, $U2$ and $U3$, we notice that the subset of structures in the basins $U1$, $U2$, $U3$, $A$, $E$, $I$,  
extracted from the clustering analysis describe all the minima extracted from the complete basis 
of the 8 free-energy profiles shown in \ref{FEP_PhiPsi}. 
More precisely, we obtain the existence of these minima but not the uniqueness of the combination of the 8 minima. 
The lack of uniqueness could have been expected as we show 
in \ref{hierarchicaldPCA} that the first 5 dPCs were usefull to describe uniquely all the different 
conformations of Met-enkephalin. The relevant point is however that the first 2 dPCs 
gives a couple of collective variables sufficient to explore all the minima, that represents a criterion to limit 
the numbers of CVs necessary to biased the metadynamics simulation.
\begin{figure*}
\includegraphics[width=1.0\textwidth, angle=0]{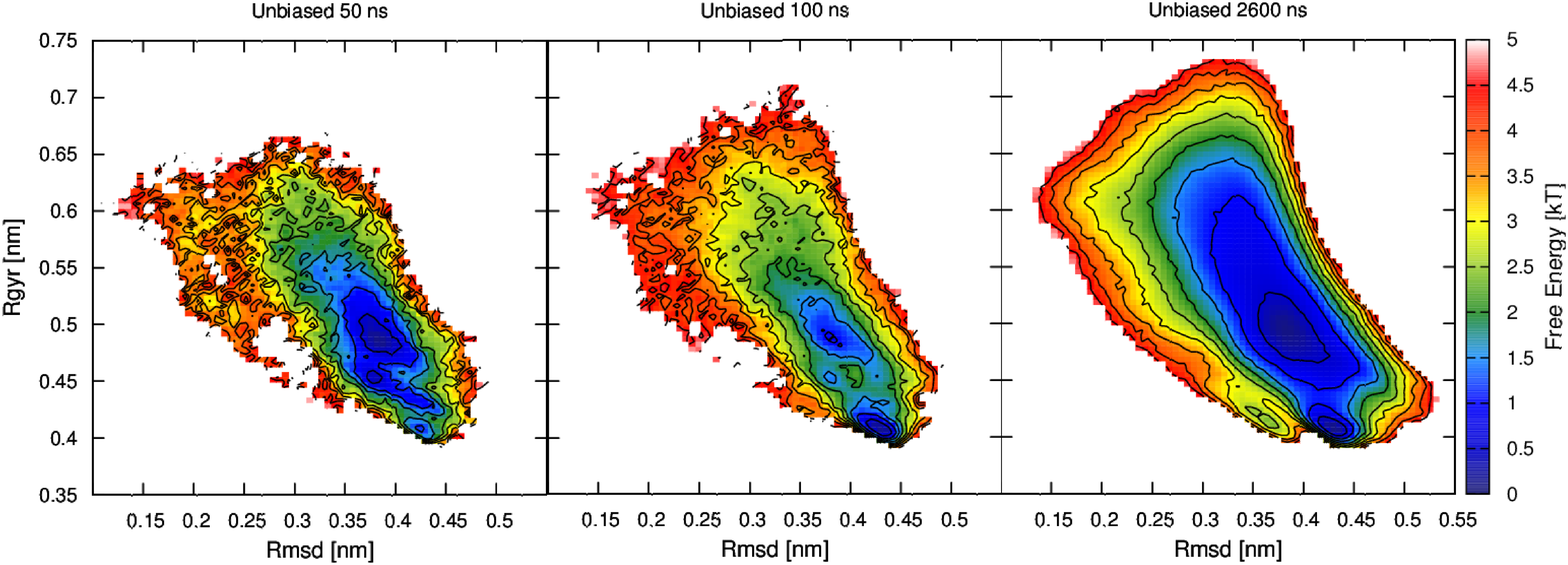}
 \vskip 0.5cm
 \includegraphics[width=1.0\textwidth, angle=0]{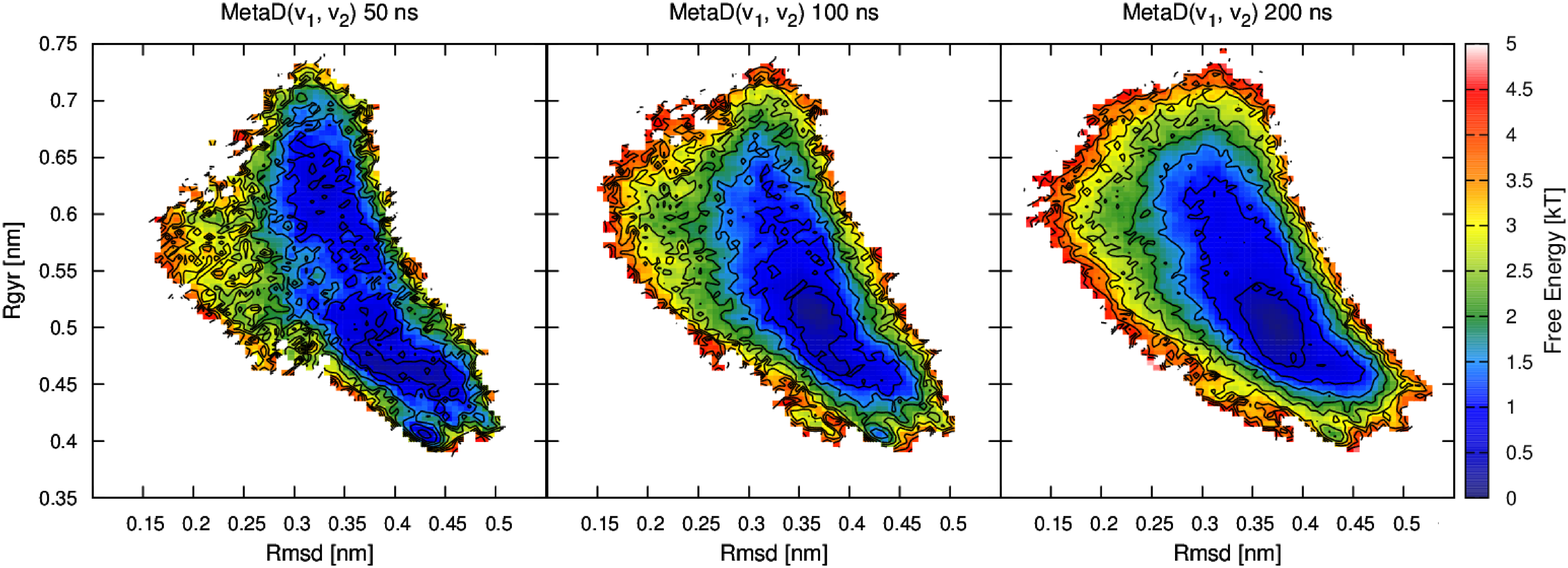}
 \vskip 0.5cm
 \includegraphics[width=1.0\textwidth, angle=0]{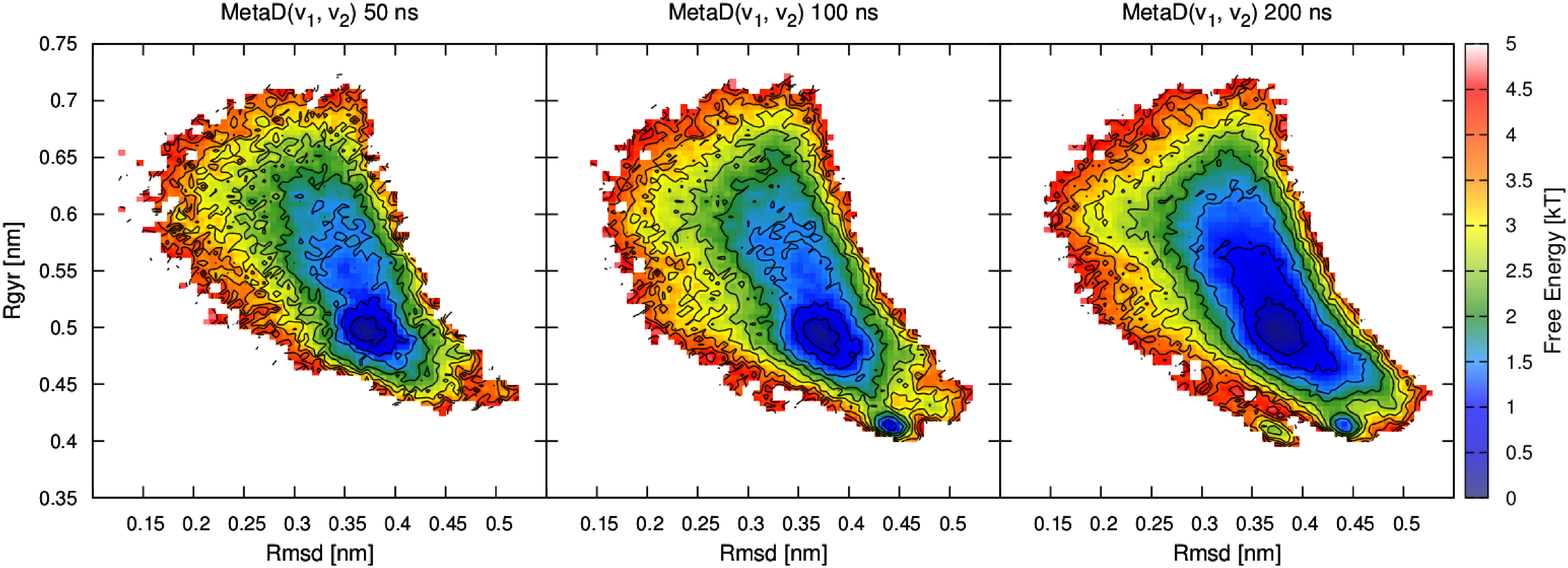}
 \caption{Free-energy surfaces for different times as a function of the root-mean-square deviation (Rmsd) 
 and the gyration radius (Rgyr) for the reference unbiased simulation (top panel) and the metadynamics simulations 
 using the two first dihedral principal components generated from the $2.6 \, \mu s$ unbiased MD trajectory (middle panel) 
 and the $26$ ns unbiased MD trajectory (bottom panel) as a bias. The contour lines are drawn every $0.5 \, k_BT$}
 \label{Fes2D_RmsdRgyr}
\end{figure*}
%

\textbf{Metadynamics reconstruction.} The metadynamics runs are performed using the first two dPCA eigenvectors 
generated from both the whole $2.6 \, \mu s$ MD trajectory and a shorter MD trajectory of $26$ ns 
 corresponding to $1\%$ of the whole unbiased MD trajectory (see Computational method). 
These collective variables take advantage of the essential dynamics technique and have been shown 
to be much more efficient than Cartesian PCA to describe the complexity of the conformational space during 
protein folding \cite{Mu-Prot2005, Maisuradze-JCTC2010}. 

The ability of these CVs to exhaustively explore the conformational space is reflected by the accuracy of the reconstructed
FES projected along two different and global observables commonly used to characterize structural properties 
of the overall conformations: the root-mean-square deviation from the initial structure (Rmsd) and the radius of gyration (Rgyr). 
We see in \ref{Fes2D_RmsdRgyr} that, to the difference with the unbiased simulation, the position of 
the minimum and the overall topology of the free-energy landscape of the metadynamics run 
performed using the dPCA eigenvectors calculated over the whole unbiased MD trajectory 
(middle panel in \ref{Fes2D_RmsdRgyr}) similarly agree with those of the reference unbiased simulation 
(upper right pannel in \ref{Fes2D_RmsdRgyr}) after the first $100$ ns. 
The same visual analysis considering the metadynamics run performed using the dPCA eigenvectors calculated over 
the $26$ ns unbiased MD trajectory (bottom panel in \ref{Fes2D_RmsdRgyr}), clearly highlights 
a slower exploration of the conformational space of Met-enkephalin in this case, as all the local minima are not explored 
after the first $100$ ns in comparison with the reference unbiased simulation (upper right pannel in \ref{Fes2D_RmsdRgyr}, 
see the region at (Rmsd, Rgyr) $\approx (0.35, \, 0.42)$).  
However, this difference could have been expected from the overlap analysis (see Computational method) as the calculated 
overlap between the dPCA eigenvectors calculated over the $26$ ns unbiased simulation and over the whole $2.6 \, \mu s$ 
simulation is $0.77$. Indeed, the metadynamics algorithm disfavors the system to explore already visited regions 
along the dPCA eigenvectors which do not correspond to the \textit{optimal} principal directions of fluctuations if 
computed over a MD trajectory of $26$ ns of duration.
\begin{figure}[th]
 \includegraphics[width=1.\columnwidth, angle=0]{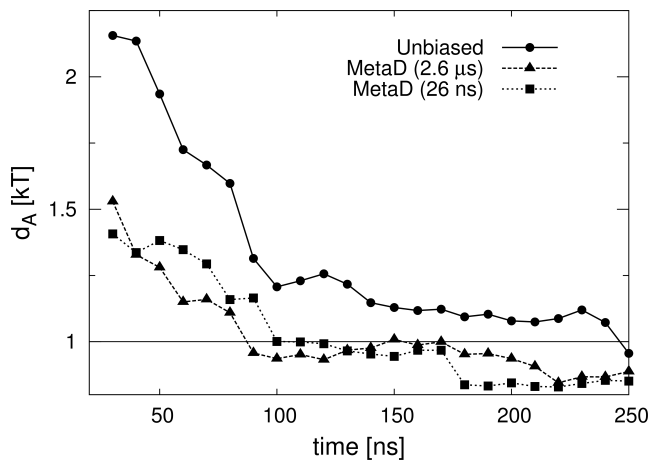}
 \caption{Comparison of the convergence of the two-dimensional FES(Rmsd,Rgyr) for the 
 unbiased and metadynamics simulations as a function of time. The similarities are calculated using as reference 
 the $2.6 \, \mu s$ unbiased simulation FES shown in \ref{Fes2D_RmsdRgyr}. The energy-function distance 
 introduced by Alonso and Echenique is considered. The result is presented in units of $k_BT$ and the 
 dashed line at 1 $k_BT$ defines the goal accuracy.}
 \label{AlonsoEchenique}
 \end{figure}
In order to asses quantitatively the convergence of the free-energy surface in the two-dimensional 
(Rmsd/Rgyr) representation, we use the correlation coefficient introduced by Alonso and Echenique \cite{Alonso-JCC2006}, 
which allows quantitative measurement of the similarity between different energy potentials (see Computational method). 
In \ref{AlonsoEchenique} we report these coefficients for the eigenvector metadynamics 
compared to the unbiased MD run as a function of time. The unbiased simulation reaches the $1\, k_BT$ reference threshold 
after $250$ ns, which is coherent with the convergence study extracted from the block analysis shown in \ref{BSE_dPCA}. 
On the other hand, the metadynamics runs (performed with the dPCA of the $26$ ns or the $2.6 \, \mu s$ 
unbiased MD trajectory) converge faster than the unbiased one, reaching and staying below the 
reference threshold after less than or approximately at $100$ ns. 
\begin{figure*}
 \includegraphics[width=0.8\textwidth, angle=0]{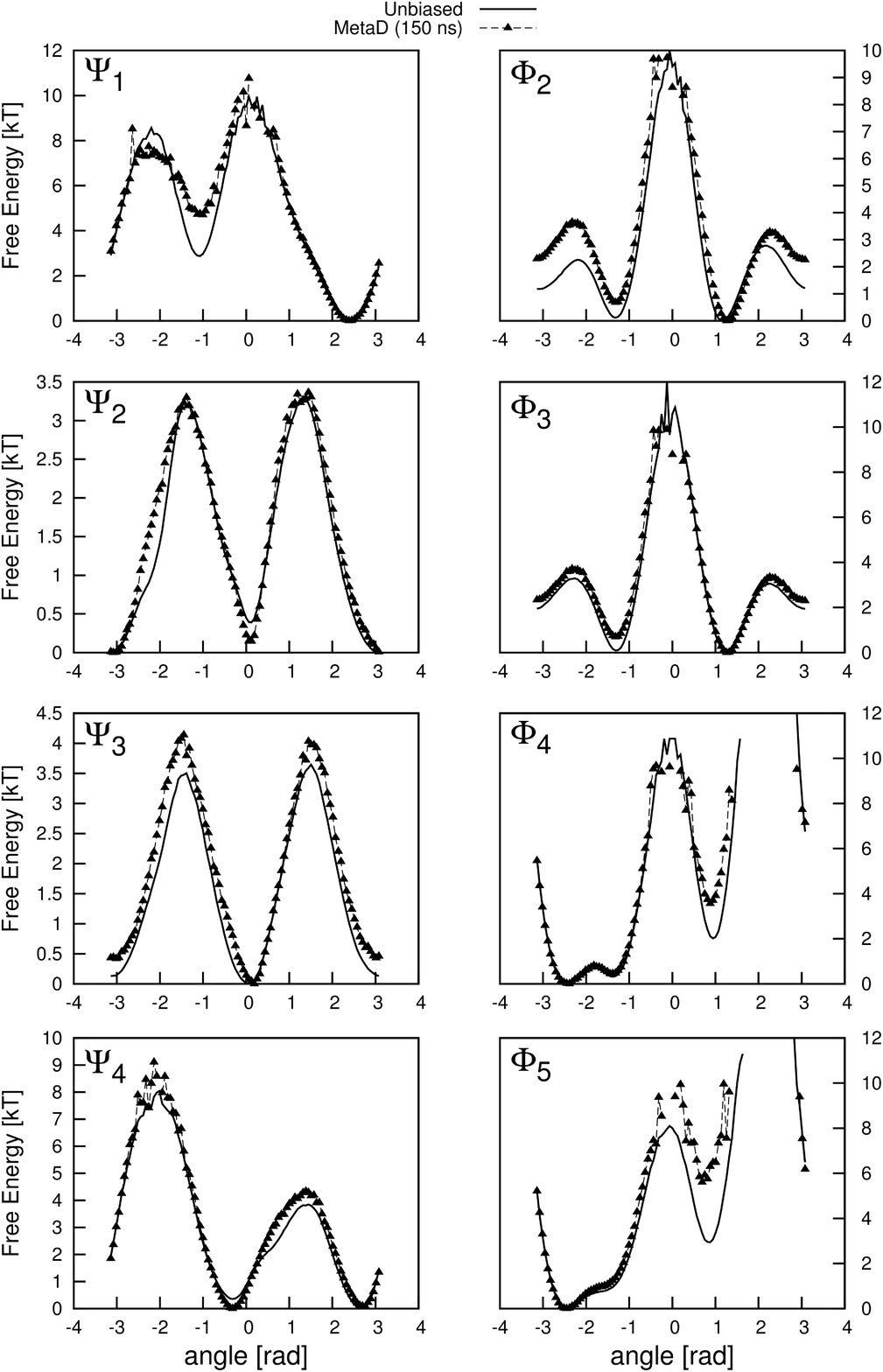}
 \caption{Representation of the eight one-dimensional free-energy profiles of the backbone angles computed from 
 the reference $2.6 \mu s$  unbiased simulation and from the first $150$ ns ($d_A < 1 \, k_BT$) of the metadynamics 
 run performed with the dPCA eigenvectors of the $26$ ns unbiased MD trajectory.}
 \label{fes1D_PhiPsi}
\end{figure*}
The convergence analysis represented in \ref{AlonsoEchenique} emphasizes the 
slower exploration of the conformational space already stated in the previous visual inspection. 
Indeed, the energy-function distance of Alonso and Echenique (cf. \ref{AlonsoEchenique}) associated to 
the WTmetaD performed with the dPCA of the $26$ ns unbiased MD trajectory reaches the reference threshold 
around $120$ ns, \textit{i.e.} sensibly after the one performed with the dPCA of the $2.6 \, \mu s$ unbiased 
MD trajectory (around $90$ ns).
\begin{figure*}
 \includegraphics[width=0.95\columnwidth, angle=0]{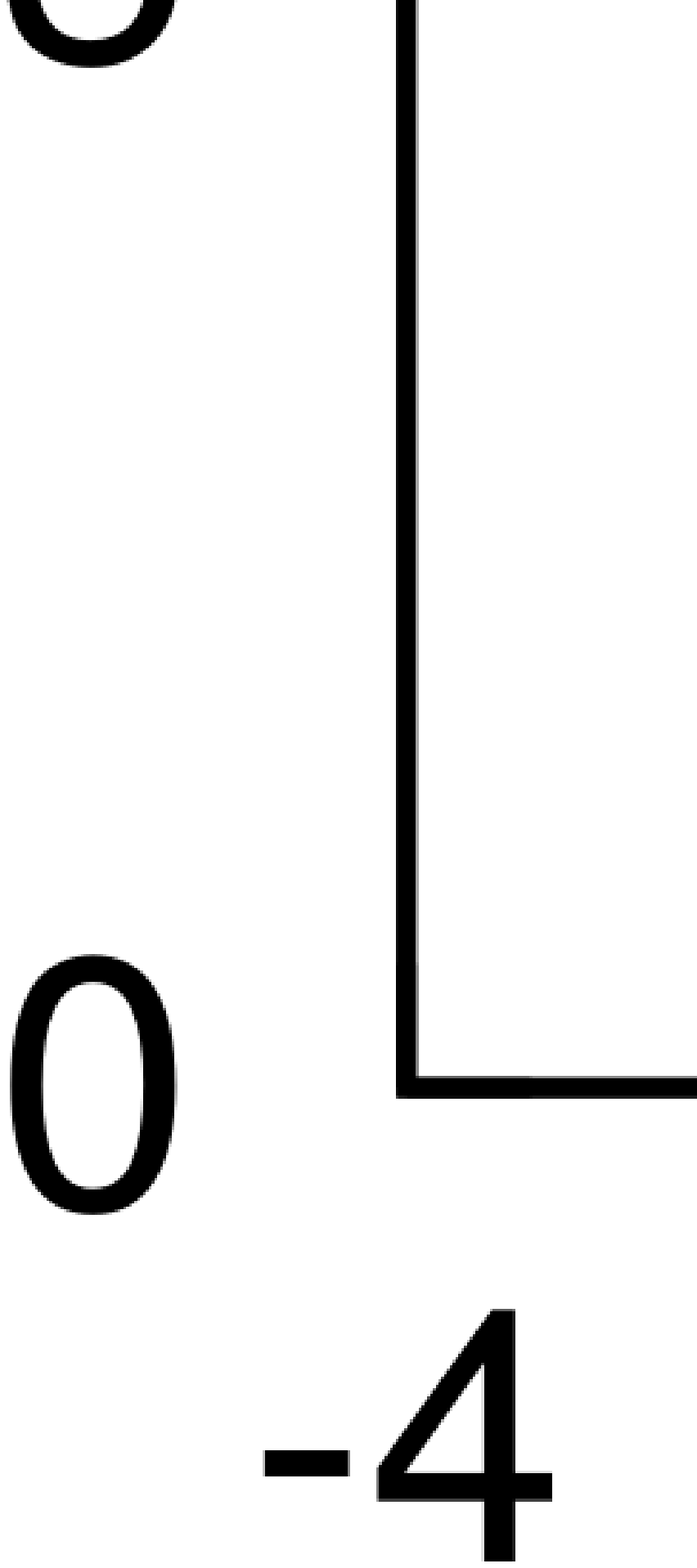}
  \hskip 1.0cm
 \includegraphics[width=0.95\columnwidth, angle=0]{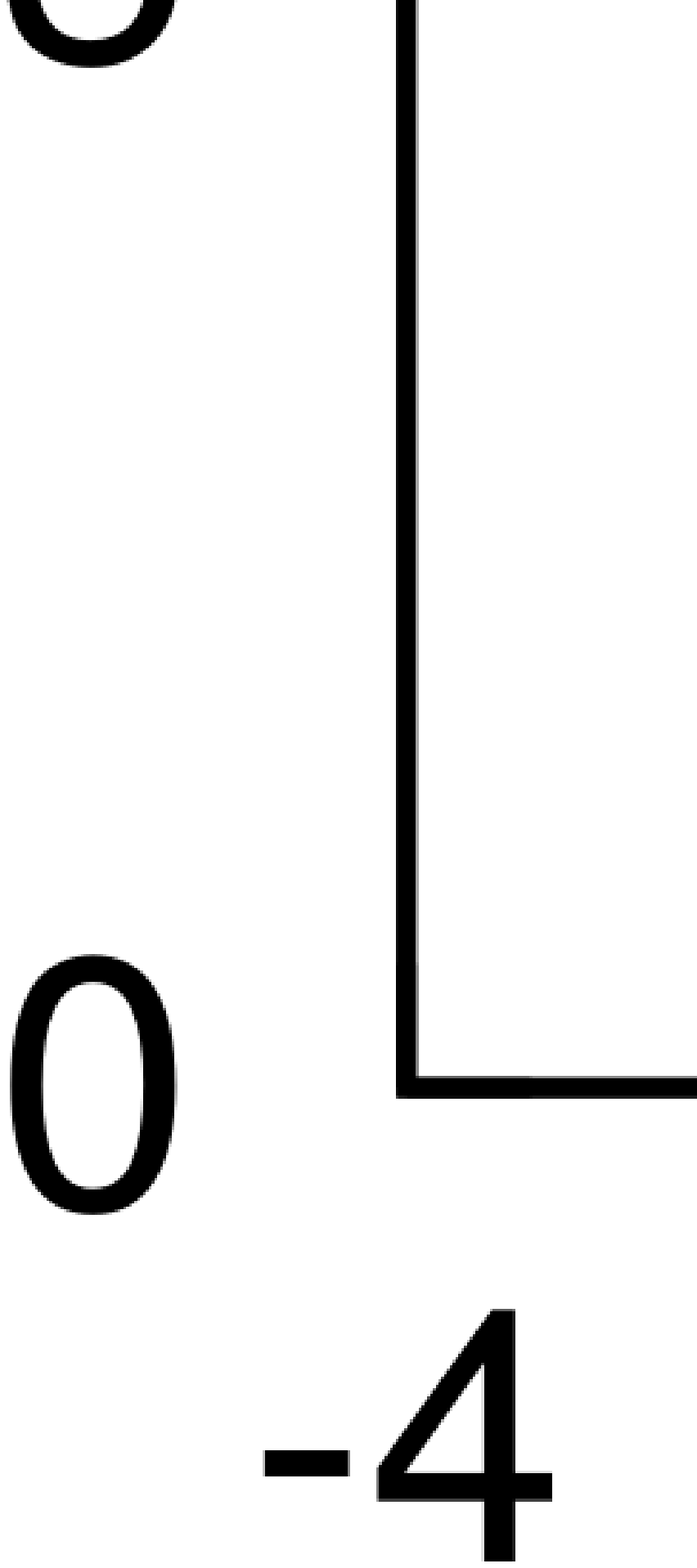}
 \caption{Comparison of the reconstruction of the FEPs associated to the Ramachandran dihedral angle $\Phi_4$ 
 for biased and unbiased simulation. The reference $2.6 \, \mu s$ unbiased simulation is represented 
 for comparison. In the left panel, we compare the WTmetaD reconstruction after $50$ ns ($d_A > 1 \, k_BT$) and 
 $150$ ns ($d_A < 1 \, k_BT$). In the right panel, the WTmetaD reconstruction at $100$ ns ($d_A < 1 \, k_BT$) 
 is compared to the unbiased one ($d_A > 1 \, k_BT$) at the same time. In both cases, we see that the biased 
 or unbiased simulation associated to the Alonso and Echenique distance $d_A > 1 \, k_BT$ completely missed 
 the local minimum on this timescale. The FEPs are shifted arbitrarily along the ordinate axis for clarity.}
 \label{comparaison_Phi4}
\end{figure*}

Let us now return to the one-dimensional FEPs of the 8 Ramachandran dihedral angles to link
the conformational space explored by the metadynamics runs with the convergence analysis based on 
the distance measure of Alonso and Echenique reported in \ref{AlonsoEchenique}. In \ref{fes1D_PhiPsi} 
it is shown the comparison between the one-dimensional FEPs of the dihedral angles for 
the reference $2.6 \,\mu s$ unbiased MD trajectory and the metadynamics run (performed using 
the dPCA eigenvectors calculated over the $26$ ns unbiased MD trajectory) after $150$ ns ($d_A < 1 \, k_BT$).
We see that the biased MD simulation captures all the different wells with very good accuracy for the 
global minima. We show for comparison in \ref{comparaison_Phi4} (left panel) that the biased MD simulation 
at $50$ ns ($d_A > 1 \, k_BT$) completely missed the local minimum associated to the Ramachandran dihedral 
angle $\Phi_4$ at $\Phi_4 \approx 1$ rad. This illustrates that the convergence criterion of the FES 
in the two-dimensional (Rmsd/Rgyr) representation can also be reformulated in terms of the capture 
of the global and local minima in the one-dimensional FEPs of the 8 Ramachandran dihedral angles.
This representation is also interesting to explain the delay of convergence of the unbiased simulation 
in comparison with the WTmetaD simulation. In \ref{comparaison_Phi4} (right panel), 
it is shown the one-dimensional 
FEP associated to the backbone angle $\Phi_4$ after $100$ ns for both the unbiased MD simulation ($d_A > 1 \, k_BT$) 
and the WTmetaD simulations ($d_A < 1 \, k_BT$) performed using the dPCA eigenvectors calculated over 
the $2.6 \, \mu$s unbiased MD trajectory. We see that in both cases, the well associated to the global minimum 
has been accuratly explored. However the unbiased simulation completely missed the local minimum 
associated to the Ramachandran dihedral angle $\Phi_4$ at $\Phi_4 \approx 1$ rad.

\section{Conclusion and Discussion}

In this paper, we considered a Well-Tempered metadynamics approach coupled with biased collective variables 
generated from a dihedral principal component analysis to reconstruct the conformational free-energy landscape 
of a small and very diffusive pentapeptide. We compared the accuracy and computational efficiency of WTmetaD 
with dPCA with a long unbiased MD simulation as well as with the previous study of Sutto et al.\cite{Sutto-JCTC2010} 
that considered WTmetaD using biased collective variables generated from Cartesian PCA. 
We reconstructed the expected free-energy surface accurately and quicker than unbiased MD simulation, 
which exhibits a quite rugged and complex appearance different from the single-minima funnel-like shape 
observed with a Cartesian PCA. 
Indeed the dPCA differs from Cartesian PCA by the fact it takes advantage of appropriate internal coordinates, 
\textit{i.e.} the Ramachandran dihedral angles describing the backbone structure of the protein. 
This approach becomes particularly usefull when one considers complex systems as multi-domain proteins \cite{Nicolai-JBSD2012} 
for which we observe diffusion motions of a domain around another one, or intrinsically disordered proteins \cite{Ozenne-JACS2012} 
characterized by a lack of stable tertiary structure due to high structural flexibility of the protein. 
In these cases, it is particularly impossible to unambiguously define a single reference structure associated 
to the Cartesian PCA procedure and necessary to eliminate the overall rotations of the molecules. 
Moreover, we underline a new criterion to limit the number of collective variables necessary to biased 
the metadynamics simulation. We consider the projection of the CVs on the complete basis 
represented by the one-dimensional FEPs of the 8 Ramachandran dihedral angles. 
The choice of the two first dPCs as CVs for the WTmetaD is then related to the capture of the different 
wells when one considers structures extracted from the clustering analysis in the free-energy map 
(dPC$^{(1)}$, dPC$^{(2)}$).
This criterion is different from the one of Sutto et al.\cite{Sutto-JCTC2010}, who justified their limitation 
to the two first PCA eigenvectors, showing that increasing this number considerably slows the convergence 
of the simulation, performing worse than the unbiased run.

Let us now discuss the efficiency of the method in the particular case of the Met-enkephalin pentapeptide.
Indeed, the dPCA over the $26$ ns, corresponding to $1\%$ of the whole unbiased MD simulation, must be linked 
to the convergence analysis of the unbiased MD simulation based either on the distance measure of Alonso 
and Echenique (cf. \ref{AlonsoEchenique}), or the block analysis (cf. \ref{BSE_dPCA}). 
We notice actually that the unbiased MD simulation converged at approximately $250$ ns, which is an order 
of magnitude smaller than the reference $2.6 \,\mu s$ unbiased MD trajectory. Thus the dPCA over the $26$ ns 
only corresponds to $10\%$ of the unbiased MD trajectory containing information about the dynamics of the system.
Nevertheless, we reconstructed the FES associated to the Met-enkephalin pentapeptide in approximately $100$ ns which is 
twice faster than the unbiased MD simulation. 
This relative modest succes of WTmetaD to accelerate the exploration of the conformational space could have been expected. 
Indeed, the small height of the free-energy barriers and diffusive behavior of the dynamics of Met-enkephalin 
is known to pose an additional challenge to free-energy-based methods like WTmetaD that generally perform better 
in the presence of high free-energy barriers and ballistic dynamics. 
Let us note that the same modest succes of WTmetaD also emerged from the 
study of Sutto et al.\cite{Sutto-JCTC2010} with the use of Cartesian PCA.
It is thus interesting to extend this method over more complex systems that present more favorable characteristics 
to apply WTmetaD. This work in progress sets out to demonstrate this approach on different well-know test proteins 
and more complex systems as multi-domain proteins that represent nowadays a real challenge for the biophysics 
community.

\section{Acknowledgment}
The authors acknowledge Y. Cote, P. Delarue and  A. Nicolai for useful discussions and advice concerning 
MD simulations with the GROMACS software package. F.S. thanks Ludovico Sutto for fruitful discussions 
concerning the PLUMED plugin for free-energy calculation. This research was conducted by using the computational 
resources of the ``Centre de Calcul de l'Universit\'e de Bourgogne'' (CCUB). 
This work was granted to the HPCresources of CINES under the allocation 2011-c2011076161 made by GENCI 
(Grand Equipement National de Calcul Intensif). F.S. thanks the Conseil Regional de Bourgogne for 
a postdoctoral fellowship (PARI-NANO2BIO).

\newpage


\begin{thebibliography}{99}
%
 \bibitem{Ansari-PNAS1985} A. Ansari, et al., Proc. Natl. Acad. Sci. \textbf{1985}, 82, 5000-5004
 \bibitem{Frauenfelder-ARBC1988} H.F. Frauenfelder, F. Parak, and R.D. Young, Annu. Rev. Biophys. Chem. \textbf{1988}, 17, 451-479
 \bibitem{Frauenfelder-Science1991} H. Frauenfelder, S.G. Sligar, and P.G. Wolynes, Science \textbf{1991}, 254, 1598-1603
 \bibitem{Onuchic-AnnuRevPhysChem1997} J.N. Onuchic, Z. Luthey-Schulten, P.G. Wolynes, Annu. Rev. Phys. Chem. \textbf{1997}, 48, 545-600 
 \bibitem{Kitao-Proteins1998} A. Kitao, S. Hayward, N. Go, Proteins \textbf{1998}, 33, 496-517
 \bibitem{Brooks-Science2001} C.L. Brooks, J.N. Onuchic, and D.J. Wales, Science \textbf{2001}, 293, 612-613
 \bibitem{Krivov-PNAS2004} S.V. Krivov, and M. Karplus, Proc. Natl. Acad. Sci. \textbf{2004}, 101, 14766-14770
 \bibitem{Wales-JPCB2006} D.J. Wales and T.V. Bogdan, J. Phys. Chem. B \textbf{2006}, 110, 20765-20776
 \bibitem{Senet-PNAS2008} P. Senet, G.G. Maisuradze, C. Foulie, P. Delarue, and H.A. Scheraga, Proc. Natl. Acad. Sci. \textbf{2008}, 105, 19708-19713
 \bibitem{Shaw-JACS2012} K. Lindorff-Larsen, N. Trbovic, P. Maragakis, S. Piana, D.E. Shaw, J. AM. Chem. Soc. \textbf{2012}, 134, 3787-3791
 \bibitem{Piana-PNAS2012} S. Piana, K. Lindorff-Larsen, and D.E. Shaw, Proc. Natl. Acad. Sci. \textbf{2012}, 109, 17845-17850
 \bibitem{Adcock-ChemRev2006} S.A. Adcock, J.A. McCammon, Chem. Rev. \textbf{2006}, 106, 1589-1615
 \bibitem{Beauchamp-JCTC2012} K.A. Beauchamp, Y.S. Lin, R. Das, V.S. Pande, J. Chem. Theory Comput. \textbf{2012}, 8, 1409-1414
 \bibitem{Bowman-JACS2011} G.R. Bowman, V.A. Voelz, and V.S. Pande, J. Am. Chem. Soc. \textbf{2011}, 133, 664-667
 \bibitem{Liwo-JPCB2007} A. Liwo, et al., J. Phys. Chem. B \textbf{2007}, 111, 260-285
 \bibitem{Kortuta-PNAS2009} A. Kortut, W.A. Hendrickson, Proc. Natl. Acad. Sci. \textbf{2009}, 106, 15667-15672
 \bibitem{Carter-CPL1989} E.A. Carter, G. Ciccotti, J.T. Hynes, R. Kapral, Chem. Phys. Lett. \textbf{1989}, 156, 472-477
 \bibitem{Bash-Science1987} P.A. Bash, U.C. Singh, F.K. Brown, R. Langridge, P.A. Kolman, Science \textbf{1987}, 235, 574-576
 \bibitem{Patey-JCP1975} G.N. Pattey, J.P. Valleau, J. Chem.Phys. \textbf{1975}, 63, 2334-2339
 \bibitem{Grubmuller-PRE1995} H. Grubm\u uller, Phys. Rev.E \textbf{1995}, 52, 2893-2906
 \bibitem{Ferrenberg-PRL1989} A. Ferrenberg, R. Swendsen, Phys. Rev. Lett. \textbf{1989}, 63, 1195-1198
 \bibitem{Jarzynski-PRL1997} C. Jarzynski, Phys. Rev. Lett. \textbf{1997}, 78, 2690-2693
 \bibitem{Darve-JCP2001} E. Darve, A. Pohorille, J. Chem. Phys. \textbf{2001}, 115, 9169-9183
 \bibitem{Gullingsrud-JCP1999} J. Gullingsrud, R. Braun, K. Schulten, J. Comput. Phys. \textbf{1999}, 151, 190-211
 \bibitem{Huber-JCAMD1994} T. Huber, A. Torda, W. van Gunsteren, J. Comput. Aided Mol. Des. \textbf{1994}, 8, 695-708
 \bibitem{Rosso-JCP2002} L. Rosso, P. Minary, Z. Zhu, M. Tuckerman, J. Comput. Phys. \textbf{2002}, 116, 4389-4402
 \bibitem{Laio-PNAS2002} A. Laio, M. Parrinello, PNAS \textbf{2002}, 99, 12562-12566
 \bibitem{Laio-RPP2008} A. Laio, F.L. Gervasio, Rep. Prog. Phys. \textbf{2008}, 71, 126601-126623
 \bibitem{Barducci-Wiley2011} A. Barducci, M. Bonomi, M. Parrinello, WIREs Comput. Mol. Sci. \textbf{2011}, 1, 826-843
 \bibitem{Barducci-PRL2008} A. Barducci, G. Bussi, M. Parrinello, Phys. Rev. Lett. \textbf{2008}, 100, 020603-020607
 \bibitem{Branduardi-JACS2005} D. Branduardi, F.L. Gervasio, A. Cavalli, M. Recantini, M. Parrinello, J. Am. Chem. Soc. \textbf{2005}, 127, 9147-9155
 \bibitem{Ichiye-Proteins1991} T. Ichiye, M. Karplus, Proteins \textbf{1991}, 11, 205-217
 \bibitem{Garcia-PRL1992} A.E. Garcia, Phys. Rev. Lett. \textbf{1992}, 68, 2696-2700
 \bibitem{Amadei-Proteins1993} A. Amadei, A.B.M. Linssen, H.J.C. Berendsen, Proteins \textbf{1993}, 17, 412-425
 \bibitem{Kitao-COSB1999} A. Kitao, N. Go, Curr. Opin. Struct. Biol. \textbf{1999}, 9, 164-169
 \bibitem{Groot-JMB2001} B.L. de Groot, X. Daura, A.E. Mark, H. Grubmuller, J. Mol. Biol. \textbf{2001}, 309, 299-313
 \bibitem{Hall-COCB2008} K.B. Hall, Curr. Opin. Chem. Biol. \textbf{2008}, 12, 612-618
 \bibitem{Daidone-Wiley2012} I. Daidone, A. Amadei, WIREs Comput. Mol. Sci. \textbf{2012}, 2, 762-770
 \bibitem{Maisuradze-Proteins2007} G.G. Maisuradze, D.M. Leitner, Proteins \textbf{2007}, 67, 569-578
 \bibitem{Maisuradze-JMB2009} G.G. Maisuradze, A. Liwo, H.A. Scheraga, J. Mol. Biol. \textbf{2009}, 385, 312-329
 \bibitem{Spiwok-JPCB2007} V. Spiwok, P. Lipovov\'a, B. Kr\'alov\'a, J. Phys. Chem. B \textbf{2007}, 111, 3073-3076
 \bibitem{Spiwok-JMM2008} V. Spiwok, B. Kr\'alov\'a, I. Tvaroska, J. Mol. Model. \textbf{2008}, 14, 995-1002
 \bibitem{Sutto-JCTC2010} L. Sutto, M.D'Abramo, F.L. Gervasio, J. Chem. Theory Comput. \textbf{2010}, 6, 3640-3646
 \bibitem{Mu-Prot2005} Y. Mu, P.H. Nguyen, G. Stock, Proteins \textbf{2005}, 58, 45-52
 \bibitem{Diamond-PS1992} R. Diamond, Protein Science, \textbf{1992}, 1, 1279-1287
 \bibitem{Altis-JCP2007} A. Altis, P.H. Nguyen, R. Hegger, G. Stock, J. Chem. Phys. \textbf{2007}, 126, 244111-244121
 \bibitem{Hughes-Nature1975} J. Hughes, T.W. Smith, H.W. Kosterlitz, L.A. Fothergill, B.A. Morgan, H.R. Morris, Nature \textbf{1975}, 258, 577-579
 \bibitem{Khaled-BBRC1976} M.A. Khaled, M.M. Long, W.D. Thompson, R.J. Bradley, G.B. Brown, D.W. Urry, Biochem. Biophys. Res. Commun. \textbf{1976}, 76, 224-232
 \bibitem{Spirtes-BBRC1978} M.A. Spirtes, R.W. Schwartz, W.L. Mattice, D.H. Coy, Biochem. Biophys. Res. Commun. \textbf{1978}, 81, 602-609
 \bibitem{Smith-Science1978} G.D. Smith, J.F. Griffin, Sciences \textbf{1978}, 199, 1214-1216
 \bibitem{Schiller-BBRC1983} P.W. Schiller, Biochem. Biophys. Res. Commun. \textbf{1983}, 114, 268-274
 \bibitem{Rapaka-ABL1985} R.S. Rapaka, V. Renugopalakrishnan, R.S. Bhatnagar, Am. Biotechnol. Lab. \textbf{1985}, 3, 11
 \bibitem{Graham-Biopoly1992} W.H. Graham, E.S. Carter, R.P. Hicks, Biopolymers \textbf{1992}, 32, 1755-1764
 \bibitem{DAlagni-EJB1996} M.D'Alagni, M. Delfini, A. Di Nola, M. Eisenberg, M. Paci, L.G. Roda, G. Veglia, Eur. J. Biochem. \textbf{1996}, 240, 540-549
 \bibitem{Sanbonmatsu-Prot2002} K.Y. Sanbonmatsu, A.E. Garcia, Proteins \textbf{2002}, 46, 225-234
 \bibitem{Shen-BJ2002} M. Shen, K. Freed, Biophys. J. \textbf{2002}, 82, 1791-1808
 \bibitem{Henin-JCTC2010} J. H\'enin, G. Fiorin, C. Chipot, M.L. Klein, J. Chem. Theory Comput. \textbf{2010}, 6, 35-47
 \bibitem{Chen-JCP2012} M. Chen, M.A. Cuendet, M.E. Tuckerman, J. Chem. Phys. \textbf{2012}, 137, 024102-024113
 \bibitem{Hess-JCTC2008} B. Hess, C. Cutzner, D. van der Spoel, E. Lindahl, J. Chem. Theory Comput. \textbf{2008}, 4, 435-447
 \bibitem{Jorgensen-JCP1983} W.L. Jorgensen, J. Chandrasekhar, J. Madura, R.W. Impey, M.L. Klein, J. Chem. Phys. \textbf{1983}, 79, 926-935
 \bibitem{Hornak-Proteins2006} V. Hornak, R. Abel, A. Okur, B. Strockbine, A. Roitberg, C. Simmerling, Proteins, \textbf{2006}, 65, 712-725
 \bibitem{Bonomi-CPC2009} M. Bonomi, D. Branduardi, G. Bussi, C. Camilloni, D. Provasi, P. Raiteri, D. Donadio, F. Marinelli, 
                          F. Pietrucci, R.A. Broglia and M. Parrinello, Comp. Phys. Comm. \textbf{2009}, 180, 1961-1972
 \bibitem{Darden-JCP1993} T. Darden, D. York, L. Pedersen, J. Chem. Phys. \textbf{1993}, 98, 10089-10092
 \bibitem{Essman-JCP1995} U. Essman, L. Perera, M.L. Berkowitz, T. Darden, H. Lee, L.G. Pedersen, J. CHem. Phys. \textbf{1995}, 103, 8577-8593
 \bibitem{Kawata-CPL2001} M. Kawata, U. Nagashima, Chem. Phys. Lett. \textbf{2001}, 340, 165-172
 \bibitem{Ramachandran-JMB1963} G.N. Ramachandran, C. Ramakrishnan, V. Sasisekharan, J. Mol. Bio. \textbf{1963}, 7, 95-99
 \bibitem{Hess-PRE2000} B. Hess, Phys. Rev. E \textbf{2000}, 62, 8438-8448
 \bibitem{Amadei-Prot1999} A. Amadei, A. Ceruso, A.M. Di Nola, Proteins: Struct. Funct. Genet. \textbf{1999}, 36, 419-424
 \bibitem{Alonso-JCC2006} J.L. Alonso, P.A. Echenique, J. Comput. Chem. \textbf{2006}, 27, 238-252
 \bibitem{Nicolai-JBSD2012} A. Nicolai, P. Delarue, P. Senet, J. Biom. Struct. Dynamics \textbf{2012}, in press
 \bibitem{Hodgkin-IJQC1987} E.E. Hodgkin, W.G. Richards,  Int. J. Quantum Chem. \textbf{1987}, 32, 105-110
 \bibitem{Grossfield-ARCC2009} A. Grossfield, D.M. Zuckerman, Annu. Rep. Comput. Chem. \textbf{2009}, 5, 23-48
 \bibitem{Hegger-PRL2007} R. Hegger, A. Altis, P.H. Nguyen, G. Stock, Phys. Rev. Lett. \textbf{2007}, 98, 0281021-0281024
 \bibitem{Cote-PNAS2012} Y. Cote, P. Delarue, P. Senet, G.G. maisuradze, and H. A. Scheraga, Proc. Natl. Acad. Sci. \textbf{2012}, 109, 10346-10351
 \bibitem{Maisuradze-JCTC2010} G.G. Maisuradze, A. Liwo, H.A. Scheraga, J. Chem. Theory Comput. \textbf{2010}, 6, 583-595
 \bibitem{Ozenne-JACS2012} V. Ozenne et al., J. Am. Chem. Soc. \textbf{2012}, 134, 15138-15148
 %
 \end{thebibliography}
\end{document}